\documentclass[journal]{IEEEtran}  
\usepackage{ifpdf}
\usepackage[dvips]{color} 
\usepackage{lmodern}

\usepackage[noadjust]{cite}

\ifCLASSINFOpdf
  \usepackage[pdftex]{graphicx}
\else
   \usepackage[dvips]{graphicx}
\fi
\usepackage{amsmath}
\usepackage{url}

\begin{document}
\title{Optical Stabilization of a Microwave Oscillator for Fountain Clock Interrogation}
%
%
%

\author{Burghard~Lipphardt, Vladislav~Gerginov, and~Stefan~Weyers
\thanks{B.~Lipphardt and S.~Weyers are with the Physikalisch-Technische Bundesanstalt (PTB), Bundesallee 100, 38116 Braunschweig, Germany (e-mail:
burghard.lipphardt@ptb.de; stefan.weyers@ptb.de).}
\thanks{V.~Gerginov was with the Physikalisch-Technische Bundesanstalt (PTB), Bundesallee 100, 38116 Braunschweig, Germany. He is now with the National Institute of Standards and Technology, 325 Broadway St, Boulder, CO 80305-3337, USA}}

%
%

\markboth{Optical Stabilization of a Microwave Oscillator for Fountain Clock Interrogation}%
{Shell \MakeLowercase{\textit{et al.}}: Bare Demo of IEEEtran.cls for IEEE Journals}
%



\maketitle

\begin{abstract}
\boldmath
We describe an optical frequency stabilization scheme of a microwave oscillator that is used for the interrogation of primary caesium fountain clocks. 
Because of its superior phase noise properties, this scheme, which is based on an ultrastable laser and a femtosecond laser frequency comb, overcomes the frequency instability limitations of fountain clocks given by the previously utilized quartz-oscillator-based frequency synthesis. The presented scheme combines the transfer of the short-term frequency instability of an optical cavity and the long-term frequency instability of a hydrogen maser to the microwave oscillator and is designed to provide continuous long-term operation for extended measurement periods of several weeks. The utilization of the twofold stabilization scheme on the one hand ensures the referencing of the fountain frequency to the hydrogen maser frequency and on the other hand results in a phase noise level of the fountain interrogation signal, which enables fountain frequency instabilities at the $2.5 \times 10^{-14} (\tau /\mathrm{s})^{-1/2}$ level that are quantum projection noise limited. 
\end{abstract}


\section{Introduction}
\label{sec_intro}

For about two decades, the most accurate realization of the SI second has been obtained from caesium fountain clocks \cite{Wynands2005}. Nowadays the best fountain clocks offer uncertainties at the low $10^{-16}$ level, while at the same time their instability allows the reaching of statistical uncertainties at the same level after averaging times of $\approx$10\,000\,s only \cite{Guena2012}. Such frequency instabilities are enabled by sufficiently increasing the signal-to-noise ratio ($\text{SNR}$) by the efficient loading of high numbers of atoms from a cold atom beam \cite{Guena2012,Dobrev2016}. Furthermore, an instability degradation due to the phase noise of the interrogation oscillator (Dick effect, \cite{Santarelli1998}) needs to be avoided. The latter has been achieved by the utilization of a cryogenic sapphire instead of a quartz-oscillator-based frequency synthesis \cite{Santarelli1999,Guena2012}. In this case, a drawback is the necessary regular costly liquid helium refill. Recently, a new generation of pulse-tube cryocoolers has been successfully tested for the microwave synthesis of fountain clocks, avoiding liquid helium refills \cite{Takamizawa2014,Abgrall2016}.

In an alternative approach, the required low oscillator phase noise level can be obtained by transferring the frequency stability of a cavity stabilized laser via a frequency comb to the microwave spectral range. For tight locking, modern combs are equipped with fast actuators (high bandwidth pump diode current controllers, intracavity electro-optic modulators) \cite{Puppe2016,Hudson2005,Zhang2012}, so that the frequency stability of the comb repetition rate and its harmonics almost reach the level of stability of the cavity stabilized laser frequency. 
Consequently, the low-noise microwave signal can be obtained directly from the femtosecond laser locked to an ultrastable optical cavity \cite{Millo2009}. Lacking the option of a fast actuator in our setup, we have chosen a differing approach by using the frequency comb as a transfer oscillator \cite{Telle2002,Lipphardt2009,Weyers2009,Tamm2014}. 
In this case, the beat frequencies between an ultrastable laser and the frequency comb, and the microwave oscillator and the comb are measured simultaneously. Since the generation of the control signal for the microwave oscillator is based on the evaluation of the frequency difference of the two beat notes, the noise contributions of the frequency comb, acting as the transfer oscillator, are suppressed within the bandwidth of the transfer. For the frequency comb stabilization, it is only required that all relevant beat frequencies are reliably kept within the bandwidths of the respective filters. 

To realize this approach, we have extended our setup for absolute frequency measurements of optical clock frequencies (e.g. \cite{Tamm2014},\cite{Huntemann2014}). We added a commercial 1.5\,$\mu$m fiber laser, locked to a high-finesse optical cavity, and transferred the laser frequency stability via the fiber laser femtosecond frequency comb to a 9.6\,GHz microwave oscillator, which is used for the synthesis of the caesium fountain interrogation signal at 9.2\,GHz. At the same time, the 9.6\,GHz microwave oscillator frequency is measured with reference to the 100\,MHz output frequency of a hydrogen maser. Based on this measurement, the long-term drift of the optical cavity and the correlated 1.5\,$\mu$m fiber laser frequency drift are compensated for by an acousto-optic modulator (AOM), which ensures the locking of the microwave oscillator to the hydrogen maser frequency in the long term.

In Section~\ref{sec_scheme} we describe our setup in detail and then present our results in Section~\ref{sec_results}.

\section{Microwave oscillator stabilization scheme}
\label{sec_scheme}

\begin{figure}[t]
\centering
  \includegraphics[width=\columnwidth]{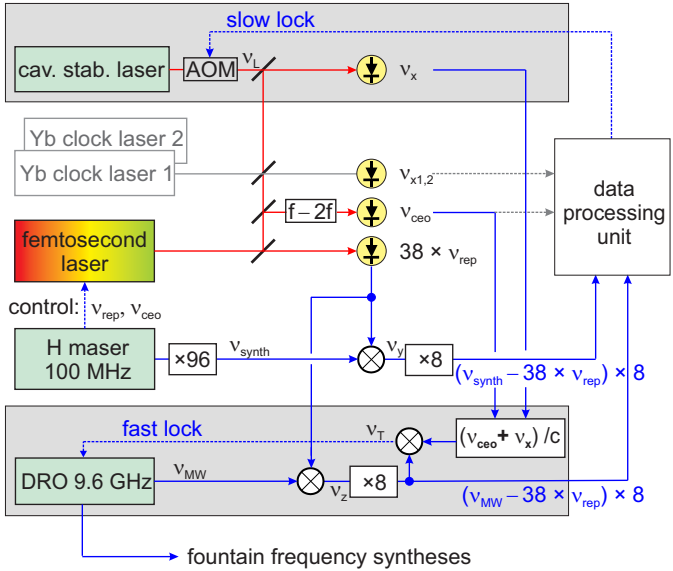}
	\caption{Setup for the generation of the optically stabilized 9.6\,GHz microwave signal and absolute optical frequency measurements. The two extra modules for microwave generation are indicated by the gray shaded boxes. DRO: dielectric resonator oscillator; AOM: acousto-optic modulator.}
	\label{fig1_Dromaser} 
\end{figure}

Our setup for absolute optical frequency measurements (Fig.~\ref{fig1_Dromaser}) has been extended by two modules (gray shaded boxes in Fig.~\ref{fig1_Dromaser}): the setup for the cavity stabilized fiber laser, and the stabilization scheme for the microwave oscillator, a dielectric resonator oscillator (DRO) at 9.6\,GHz. The core of the whole setup is a frequency comb (FC1500 from Menlo Systems), based on an erbium-doped fiber laser, with a repetition rate $\nu_\mathrm{rep}$ of about 252\,MHz at a wavelength of 1.5\,$\mu$m. The oscillator of the frequency comb generates pulses of 100\,fs in length. The corresponding optical comb has a Fourier limited spectrum width of 30\,nm. In principle, optical detection of the 100\,fs laser pulses generates a ``second frequency comb'' consisting of the comb repetition rate and its harmonics extending well into the microwave spectral range up to 4\,THz, given by the Fourier limit. Absolute frequency measurements of optical frequencies and the transfer of stability of optical frequencies into the microwave regime are enabled, because of the phase coherence between the optical comb frequencies and the comb repetition rate.

To fulfill the requirements of our microwave oscillator stabilization scheme regarding optical power and bandwidth, we have chosen a fast InGaAs-PIN photodiode (DSC 40S, BW\,=\,12\,GHz) for the detection of the repetition rate and its harmonics. The 38$^{\mathrm{th}}$ harmonic (at $\approx 9.6$\,GHz) is subsequently selected by a band-pass filter (bandwidth 400\,MHz) and amplified by a microwave amplifier with phase noise specification (HMC-C050).  

The detection and amplification constitute a fundamental limit for the short-term frequency stability transfer from the optical to the microwave domain in our setup. An optical power of 0.5\,mW at the photodiode results in $-27$\,dBm electrical power of the 38$^\mathrm{th}$ harmonic and thus in a calculated $\text{SNR}$ of 144\,dB/Hz. The amplifier limits the $\text{SNR}$ to 140\,dB/Hz close to the carrier at 100\,Hz and 142\,dB/Hz at frequencies larger than 1\,kHz. 

One part of the 38$^{\mathrm{th}}$ harmonic's signal of the repetition rate is used for the fast locking of the microwave oscillator (see below). The other part of the signal is mixed with a 9.6\,GHz frequency synthesis output signal $\nu_\mathrm{synth}$, referenced to the maser. The resulting beat signal $\nu_\mathrm{y}=38\times \nu_\mathrm{rep}-\nu_\mathrm{synth}$ is kept constant via an offset lock for locking the repetition rate of the frequency comb to the 9.6\,GHz signal with a bandwidth of about 1\,kHz (not shown in Fig.~\ref{fig1_Dromaser}). 

Three erbium-doped fiber amplifiers (not shown in Fig.~\ref{fig1_Dromaser}), including matched nonlinear optical components, are employed to spread and optimize the femtosecond laser spectrum. One of these amplifiers generates the octave-spanning spectral range of 1--2\,$\mu$m to provide the carrier envelope offset frequency $\nu_\mathrm{ceo}$ of the frequency comb, which is realized in a $f-2f$ scheme \cite{Jones2000, Holzwarth2000}. The offset frequency ($\text{SNR}$\,=\,42\,dB at RBW\,=\,300\,kHz) is locked to an intermediate frequency of 20\,MHz with a bandwidth of a few kHz (not shown in Fig.~\ref{fig1_Dromaser}). The spectrum core areas of the two other amplifiers are tuned to optimize the beat notes $\nu_\mathrm{x1},\nu_\mathrm{x2}$ with the optical clock transition frequencies of our quadrupole ({\em E}\/2) and octupole ({\em E}\/3) $^{171}$Yb$^+$ frequency standards \cite{Tamm2014}, \cite{Huntemann2016}. For optical frequency measurements, the  frequencies $\nu_\mathrm{rep}$, $\nu_\mathrm{ceo}$ and the difference frequencies $\nu_\mathrm{x1},\nu_\mathrm{x2}$ between the frequencies of the optical frequency standards and the individual comb teeth are registered by synchronous multichannel counters (K+K FXE \cite{Kramer2004}), free of dead times. From the counter results, absolute optical frequencies, which are referenced to the maser and the caesium fountains, are calculated in a data processing unit (Fig.~\ref{fig1_Dromaser}).

To provide the short-term stability of the optically stabilized DRO, a commercial fiber laser (Koheras BASIK E15) at a wavelength of 1.54\,$\mu$m is locked to an optical cavity made of ultra low expansion (ULE) glass with highly reflecting ULE mirrors by means of the Pound-Drever-Hall technique. With a cavity length of 75\,mm (FSR\,=\,2\,GHz), a finesse of 320\,000 is obtained. The control bandwidth of this servo loop is limited by an AOM to 100\,kHz. Low-frequency variations are compensated for by changing the effective fiber length of the laser using a piezo element adjusted by an integral element in the control loop. As a result, a frequency reference with a short-term stability of $10^{-15}$ for averaging times between 1 and 10\,s is obtained with a prevailing linear drift of 40\,mHz/s. 

To realize the transfer oscillator concept \cite{Telle2002}, we need to simultaneously track the optical beat signals $\nu_\mathrm{x}$ and $\nu_\mathrm{ceo}$ and the microwave beat signal $\nu_\mathrm{z}$ between the DRO and the frequency comb. While the optical and microwave beat notes are from different spectral regions, they are based on strongly phase coupled modes of the same oscillator. Processing of the beat notes enables their subtraction from each other, which effectively eliminates the noise of the frequency comb. 

The cavity stabilized laser light and the light from the frequency comb are combined via single-mode fibers. The beat frequency $\nu_\mathrm{x}$ (27\,MHz, $\text{SNR}$\,=\,43\,dB at RBW\,=\,300\,kHz) from the laser light (frequency $\nu_\mathrm{L}$) and the $m^\mathrm{th}$ mode of the frequency comb ($m=769\,794$) is detected by a standard InGaAs photodiode. An absolute frequency measurement with reference to the hydrogen maser frequency is reduced to the counting of the three radio frequencies $\nu_\mathrm{rep}$, $\nu_\mathrm{ceo}$ and $\nu_\mathrm{x}$:

\begin{equation}
\nu_\mathrm{L}=m \times \nu_\mathrm{rep}  + \nu_\mathrm{ceo} + \nu_\mathrm{x}. 
\label{laserf}
\end{equation}

The DRO frequency $\nu_\mathrm{MW}$ is mixed with the 38$^\mathrm{th}$ harmonic of the repetition rate, which yields $\nu_\mathrm{MW}=38 \times \nu_\mathrm{rep} +\nu_\mathrm{z}$. Eliminating $\nu_\mathrm{rep}$ by utilizing (\ref{laserf}), we obtain the microwave frequency $\nu_\mathrm{MW}$ as

\begin{equation}
\nu_\mathrm{MW}=\frac{38}{m} \times \nu_\mathrm{L} + [\nu_\mathrm{z}-\frac{38}{m} (\nu_\mathrm{ceo}+\nu_\mathrm{x})]
\label{MW}
\end{equation}

\noindent which establishes a fixed relation between $\nu_\mathrm{MW}$ and $\nu_\mathrm{L}$, if the term within the square brackets is kept constant. The latter contains beat frequencies generated in the microwave ($\nu_\mathrm{z}$\,=\,5.9\,MHz) and in the optical spectral region ($\nu_\mathrm{ceo}+\nu_\mathrm{x}$\,=\,47\,MHz), which are separated by a large scaling factor of $m/38$.  

The spectral range adaptation of the beat notes is realized by a multiplication of $\nu_\mathrm{z}$ and a division of $\nu_\mathrm{ceo}+\nu_\mathrm{x}$. For this purpose, an extra module, an analogue data processor with a tracking bandwidth of about 1\,MHz, is set up: An overtone oscillator generates the 8$^\mathrm{th}$ harmonic of $\nu_\mathrm{z}$, which reduces the further requirements of the $\text{SNR}$ and also of the counter resolution for data processing. Furthermore, $\nu_\mathrm{ceo}$ and $\nu_\mathrm{x}$ are added in a mixer, filtered by a tracking oscillator, amplified and divided by $c=m/(38 \times 8)$.
The division is undertaken in a four-step process: In three identical stages 100\,MHz is added each time, before the sum frequency is divided by a factor of 8. Before the last of these stages another stage is inserted, in which 100\,MHz is also added, before the sum frequency is divided by a scaling factor of 4.9457\ldots which is obtained from a direct digital synthesizer with a 48-bit accumulator (AD9956). 

The output frequency of the division process ($\approx$15\,MHz) is mixed with the overtone frequency of $\nu_\mathrm{z}$. As a final result, we obtain a transfer beat frequency $\nu_\mathrm{T}$ equal to the eightfold of the term in square brackets of (\ref{MW}):

\begin{equation}
\nu_\mathrm{T}=\nu_\mathrm{z} \times 8 -\frac{\nu_\mathrm{ceo}+\nu_\mathrm{x}}{c} = \nu_\mathrm{MW} \times 8 - \frac{\nu_\mathrm{L}}{c}. 
\end{equation}

\noindent Since $\nu_\mathrm{T}$ is independent of $\nu_\mathrm{rep}$ and $\nu_\mathrm{ceo}$, fluctuations of these frequencies are effectively suppressed, and $\nu_\mathrm{T}$ comprises the direct phase comparison between the DRO and the stabilized fiber laser. The comparison result is used to control the DRO with a bandwidth of $\approx$50\,kHz using an offset lock (``fast lock'' in Fig.~\ref{fig1_Dromaser}).

Finally, to ensure the locking of the DRO to the maser, we need to compensate for the frequency drift of the optical cavity by locking $\nu_\mathrm{L}$ to the maser. For this purpose, we employ a slow lock (time constant $\approx$50\,s), which is based on a comparison of the maser frequency and the DRO frequency in the data processing unit, using an AOM for the adjustment of $\nu_\mathrm{L}$ (see Fig.~\ref{fig1_Dromaser}).

The utilized ``state-of-the-art'' DRO (PSI, $-$\,110\,dBc at 10\,kHz) has a tuning bandwidth of a few 100\,kHz. The 100\,MHz hydrogen maser reference frequency is delivered to our setup via a semi-rigid coaxial cable by a commercial distribution amplifier (SDI…) with galvanic isolation. All synthesizers and counters of the setup are referenced to the maser frequency. A homemade frequency synthesis \cite{Gupta2007} is utilized to generate the 96$^\mathrm{th}$ harmonic $\nu_\mathrm{synth}$ of the 100\,MHz signal. 

The 9.6\,GHz optically stabilized microwave signal from the maser-referenced DRO (Fig.~\ref{fig1_Dromaser}) is split, amplified and delivered to the two syntheses of the PTB fountain clocks CSF1 \cite{Weyers2001,Weyers2001b} and CSF2 \cite{Gerginov2010,Weyers2012} to generate the 9.193\,GHz signal needed for the atom interrogation. The syntheses are provided with electronic switches which enable either the utilization of the optically stabilized 9.6\,GHz signal or alternatively a 9.6\,GHz signal from another DRO stabilized to a low-noise quartz oscillator \cite{Gupta2007}. In the case of a failure of the optically stabilized microwave signal, the fountain control software detects the resultant exceeding of the limits for the measured atom number, the transition probability or the frequency deviations, and arranges the automatic switching to the quartz-based 9.6\,GHz signal.

\section{Results}
\label{sec_results}

\subsection{Characterization of the optically stabilized microwave oscillator signal}
\label{sec_CharDdro}

\begin{figure}
\centering
  \includegraphics[width=\columnwidth]{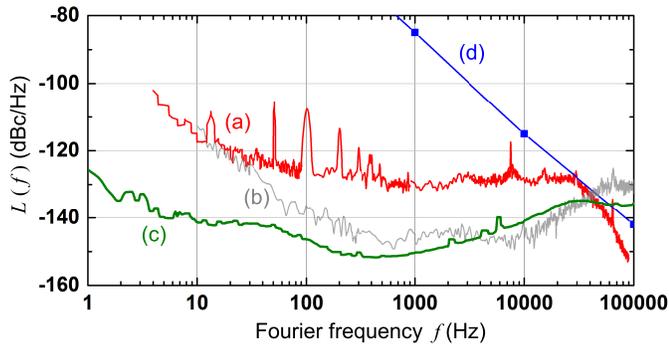}
	\caption{Single-sideband phase noise power spectral density $L$ relative to the carrier at 9.6\,GHz. (a) red: measured phase noise spectrum of the employed dielectric resonator oscillator (DRO) for the case that it is locked to an ultrastable laser using the transfer oscillator scheme. (b) gray: measured phase noise spectrum of the transfer beat frequency signal $\nu_\mathrm{T}$. (c) green: measured phase noise spectrum of the transfer beat frequency signal $\nu_\mathrm{T}$ after modification of the offset lock of the transfer beat (see text). (d) blue: specified phase noise of the unstabilized DRO.}
	\label{fig2_Pn} 
\end{figure}

For the characterization of the optically stabilized microwave signal, a second frequency comb independently stabilized to the same laser was utilized. The short-term characterization by a phase noise measurement yielded a phase noise level of $-$117\,dBc/Hz at 10\,Hz from the carrier frequency [curve (a) in Fig.~\ref{fig2_Pn}] \cite{Tamm2014}. Noise integration up to 30\,kHz results in an Allan frequency deviation of $4 \times 10^{-15}$ (1\,s), dominated by the increase of the noise level at Fourier frequencies below 100\,Hz. Also the noise analysis of the transfer beat frequency signal $\nu_\mathrm{T}$ (in-loop) showed the same limitation [curve (b) in Fig.~\ref{fig2_Pn}], which was given by the noise properties of the utilized synthesizer for the offset lock to keep the transfer beat frequency $\nu_\mathrm{T}$ constant along with the limited selectivity of the utilized spectrum analyzer. 

Recently, the limitation has been overcome by a modification of the offset lock of the transfer beat avoiding additional noise from the offset synthesizer. Moreover, the servo loop bandwidth of the fast lock was increased to more than 100\,kHz by utilizing filters with smaller delays. The resulting transfer beat spectrum is depicted in Fig.~\ref{fig2_Pn} [curve (c)]. The expectation that the previous phase noise limitation at low Fourier frequencies has been overcome is supported by a frequency ratio measurement of the DRO and the clock laser of the $^{171}$Yb$^+$ single-ion frequency standard with the same frequency comb, which exhibited a significant improvement of the frequency instability. However, a real out-of-loop noise measurement would be needed to definitely confirm the removal of the former limitation.

\begin{figure}
\centering
  \includegraphics[width=\columnwidth]{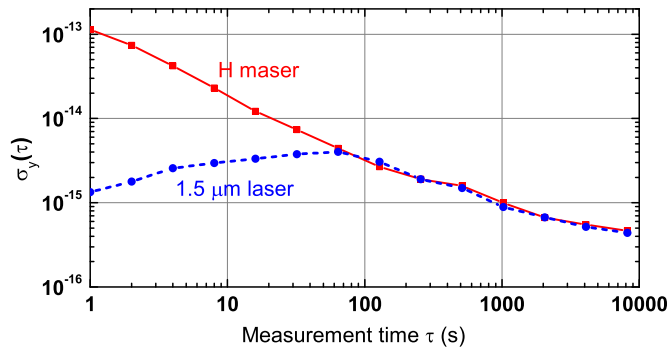}
	\caption{Measured Allan standard deviations $\sigma_y (\tau)$ of the frequencies of the hydrogen maser (solid line, red) and the cavity stabilized 1.5\,$\mu$m fiber laser (dashed line, blue), locked to the hydrogen maser. For both measurements, the output frequency of the $^{171}$Yb$^+$ single-ion frequency standard served as a reference, which only contributes at a negligible level to the measured standard deviations.}
	\label{fig3_DROADEV} 
\end{figure}

The long-term performance of the optically stabilized microwave signal is characterized in terms of the Allan standard deviation. We measured the frequency instability of the hydrogen maser and of the cavity stabilized fiber laser locked to the hydrogen maser (Fig.~\ref{fig3_DROADEV}). In both cases, the output frequency of the $^{171}$Yb$^+$ frequency standard served as a reference, which only contributes at a negligible level to the measured standard deviations. In the long term ($\tau >$~100\,s), both Allan deviations agree very well. Since the DRO is locked to the cavity stabilized fiber laser with a bandwidth of 100\,kHz, the microwave signal exhibits the same long-term behavior. 

If required, the short-term stability of the microwave signal could be further improved: On the one hand, superior radio-frequency amplifiers with lower phase noise could be utilized. Secondly, the amplitude to phase conversion in the fast InGaAs-PIN photodiode for the detection of the repetition rate and its harmonics can be reduced by stabilizing the optical power \cite{Taylor2011,Zhang2012a} and selecting the operating point of the photodiode \cite{Zhang2010}. Finally, the $\text{SNR}$ can be increased by ``multiplication of the pulse rate'' \cite{Haboucha2011}.

\subsection{Resulting frequency instabilities of the fountain clocks CSF1 and CSF2}
\label{sec_InstabCSF}

In the case of the operation of a fountain clock which is quantum projection noise limited, the $\text{SNR}$ is given by $\text{SNR}=N_{\mathrm{\mathrm{at}}}^{1/2}$, with $N_{\mathrm{\mathrm{at}}}$ the total detected number of atoms in the F\,=\,3 and F\,=\,4 hyperfine components of the Cs ground state \cite{Santarelli1999}. The frequency instability expressed by the Allan deviation is given by: 

\begin{equation}
\label{sigmay}
	\sigma_y\left(\tau\right)=\frac{1}{\pi}\frac{\Delta\nu}{\nu_0}\frac{1}{\text{SNR}}\sqrt{\frac{T_c}{\tau}}
\end{equation}

\noindent where $\Delta\nu$ is the full-width-at-half-maximum of the Ramsey fringe, $\nu_0$\,=\,9\,192\,631\,770\,Hz is the clock transition frequency, $T_c$ is the cycle time, and $\tau$ the measurement time. The precondition is that an instability degradation due to the phase noise of the interrogation oscillator (Dick effect, \cite{Santarelli1998}) can be avoided. The instability described by (\ref{sigmay}) is reduced with increased detected atom numbers (from increased loading times) as long as the factor with the square root of $T_c$ (also containing the loading time) does not become too large. This consideration gives a first criterion for the best choice of the employed loading time (and for the resulting detected atom number $N_{\mathrm{\mathrm{at}}}$ and cycle time $T_c$). To achieve the best compromise for the combined statistical and systematic uncertainties, one has to furthermore take into account that in general the systematic part of the uncertainty due to cold atom collisions scales with $N_{\mathrm{\mathrm{at}}}$ (e.g. see \cite{Gerginov2010}).  

In the fountain CSF1, where magneto-optical trap (MOT) loading is utilized, the collisional frequency shift is evaluated by simply varying the microwave field power in the state selection cavity to operate the fountain at different atomic densities. This gives a 10\% systematic uncertainty estimate for the collisional shift \cite{Weyers2001b}. As a result, the usable $N_{\mathrm{\mathrm{at}}}$ is limited to keep the related uncertainty contribution at a reasonable level. We use an MOT loading time of 163.2\,ms ($T_c$\,=\,1.1145\,s), which optimizes the overall uncertainty. In the fountain CSF2 with molasses loading, the method of rapid adiabatic passage is employed for the collisional shift determination \cite{Kazda2013}. Since the systematic uncertainty of the collisional shift is below 0.5\%, much higher atom numbers are compatible with systematic collisional shift uncertainties below the $10^{-16}$ level. Molasses loading from a cold atom beam \cite{Dobrev2016} enables a relatively short loading time of 340\,ms ($T_c$\,=\,1.2345\,s) for the minimization of the overall uncertainty. Increasing the loading time to 690\,ms ($T_c$\,=\,1.5845\,s), however, optimizes the statistical uncertainty.

With both fountains, $\text{SNR}$ measurements utilizing either two $\pi/2$ pulses (at frequency detuning of half of the full linewidth) or two $\pi/4$ pulses (at zero frequency detuning) yield the same $\text{SNR}$. Since the $\text{SNR}$ is sensitive to frequency noise only in the first case, this finding is the first proof that the noise contribution by the optically stabilized microwave signal is negligible. The measured $\text{SNR}$ values are given in Table~\ref{table_instabilities} together with the resultant calculated frequency instabilities $\sigma_\mathrm{y} (\mathrm{1\,s})$ using (\ref{sigmay}).

\begin{table*}[t]
\renewcommand{\arraystretch}{1.0}
\caption{Loading and Cycle Times, Signal-to-Noise Ratios and Frequency Instabilities of the Fountain Clocks CSF1 and CSF2}
\label{table_instabilities}
\centering
\begin{tabular}{lcccccc}
\hline
\rule[-3mm]{0mm}{8mm} & Loading Time & Cycle Time & $\text{SNR}$ & $\sigma_\mathrm{y} (\mathrm{1\,s})$ & $\sigma_\mathrm{y, DRO} (\mathrm{1\,s})$ & $\sigma_\mathrm{y, measured} (\mathrm{1\,s})$\\
\hline\hline
\rule[-3mm]{0mm}{8mm}\bfseries CSF1 & 163.2\,ms & 1.1145\,s & 350 & $9.2 \times 10^{-14}$ & $4.8 \times 10^{-15}$ & $9.1 \times 10^{-14}$\\
\hline
\rule[-3mm]{0mm}{8mm}\bfseries CSF2 & 340\,ms & 1.2345\,s & 1060 & $3.3 \times 10^{-14}$ & $5.7 \times 10^{-15}$ & $3.4 \times 10^{-14}$\\
\rule[-3mm]{0mm}{5 mm}& 690\,ms & 1.5845\,s & 1660 & $2.4 \times 10^{-14}$ & $7.8 \times 10^{-15}$ & $2.5 \times 10^{-14}$\\
\hline
\multicolumn{7}{l}{The frequency instabilities $\sigma_\mathrm{y}$ scale with $(\tau/1\,\mathrm{s})^{-1/2}$. The instability $\sigma_\mathrm{y} (\mathrm{1\,s})$ is calculated from the}\rule[0mm]{0mm}{4mm}\\
\multicolumn{7}{l}{measured $\text{SNR}$ (see text) by using (\ref{sigmay}). The instability contribution $\sigma_\mathrm{y, DRO} (\mathrm{1\,s})$ is caused by the Dick}\\
\multicolumn{7}{l}{effect \cite{Santarelli1998} and calculated from the measured DRO phase noise (Fig.~\ref{fig2_Pn}).}\\
\end{tabular}
\end{table*}

Taking into account the measured phase noise level of the optically stabilized DRO (Fig.~\ref{fig2_Pn}), we calculated the frequency instability contribution caused by the Dick effect \cite{Santarelli1998} for the different cycle times of the two fountain clocks ($\sigma_\mathrm{y, DRO} (\mathrm{1\,s})$ in Table~\ref{table_instabilities}). It becomes obvious that even with the former measured phase noise level shown in Fig.~\ref{fig2_Pn}, which is a worst case estimate as explained in Section~\ref{sec_CharDdro}, the resulting instability contribution of the optically stabilized DRO is negligible compared to the calculated instabilities $\sigma_\mathrm{y} (\mathrm{1\,s})$.

\begin{figure}
\centering
  \includegraphics[width=\columnwidth]{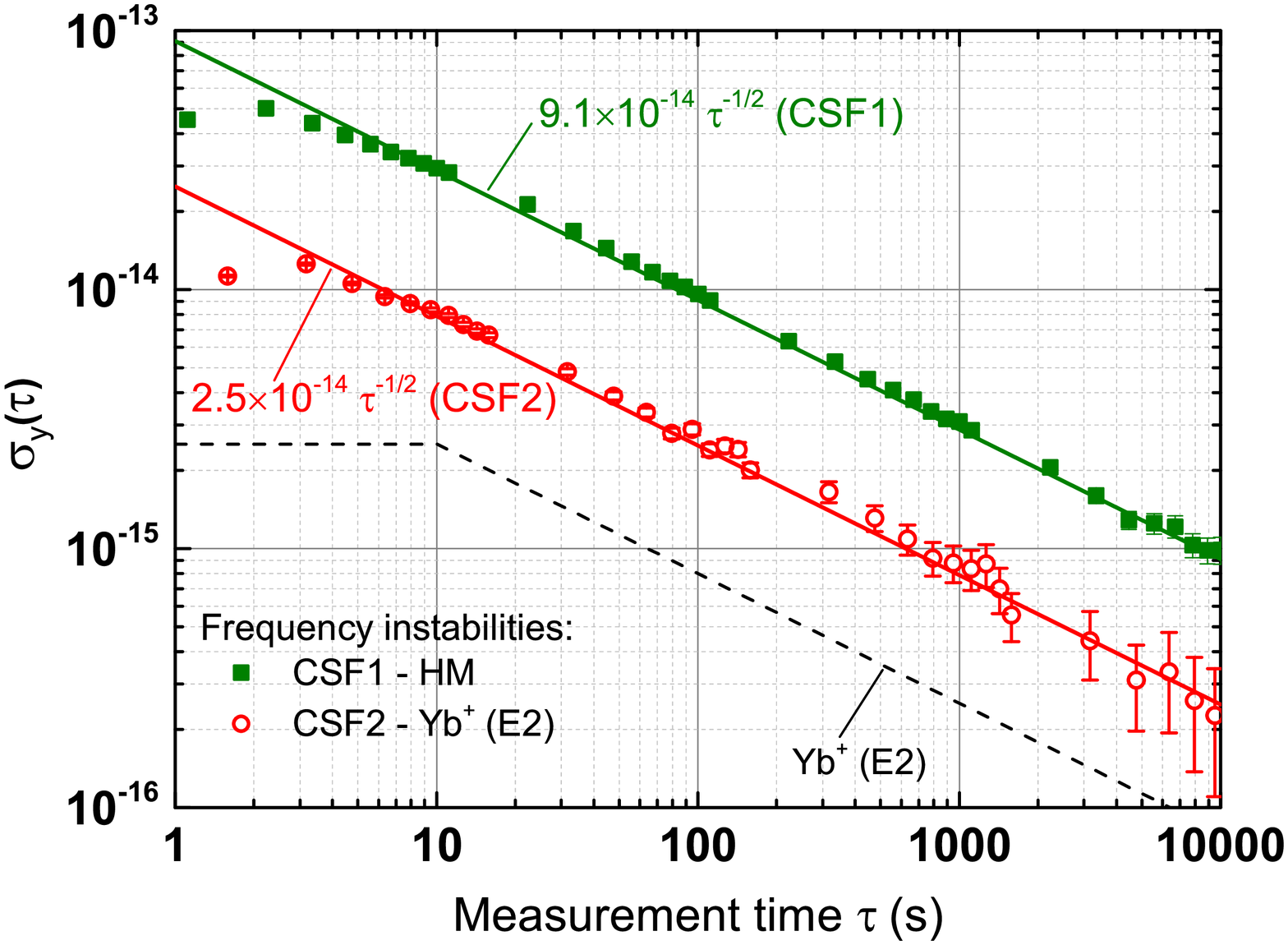}
	\caption{Measured Allan standard deviations $\sigma_y (\tau)$ of the CSF1 [full data points (green)] and CSF2 [open data points (red)] frequencies. For the CSF1 measurement, a hydrogen maser served as the reference while the CSF2 frequency was referenced to the output frequency of the  $^{171}$Yb$^+$ single-ion frequency standard [quadrupole (E2) transition], whose instability level is indicated by the dashed line. Subtraction of the noise levels of the maser and the single-ion frequency standard, respectively, yields the instability levels indicated by the full lines (CSF1: $9.1 \times 10^{-14} (\tau/1\,\mathrm{s})^{-1/2}$, CSF2: $2.5 \times 10^{-14} (\tau/1\,\mathrm{s})^{-1/2}$).}
	\label{fig4_CSFADEV} 
\end{figure}

In Fig.~\ref{fig4_CSFADEV}, the measured Allan standard deviations for fountain frequency measurements with cycle times $T_c$\,=\,1.1145\,s (CSF1) and $T_c$\,=\,1.5845\,s (CSF2) are presented. The results $\sigma_\mathrm{y, measured} (\tau)=9.1 \times 10^{-14} (\tau/1\,\mathrm{s})^{-1/2}$ for CSF1 and $\sigma_\mathrm{y, measured} (\tau)=2.5 \times 10^{-14} (\tau/1\,\mathrm{s})^{-1/2}$ for CSF2 agree very well with the instabilities calculated from the $\text{SNR}$ data (see Table~\ref{table_instabilities}) and prove again that the instability contributions of the optically stabilized DRO are negligible at the current fountain performance levels.

\subsection{Operation of the optically stabilized microwave oscillator signal}
\label{sec_FrequMeas}

In March~2014, the first continuous 10-day measurement (dead time $\approx $1\%) with the fountains and the presented optically stabilized DRO scheme was performed to contribute to the monthly calibration of International Atomic Time (TAI) by the Bureau International des Poids et Mesures (BIPM). Both fountain frequencies agreed at the $10^{-16}$ level, which was very well within their combined uncertainty. 

Since June~2015, the optically stabilized DRO scheme has been routinely used for fountain operation. A number of TAI calibrations as well as optical frequency measurements have been performed with both fountains. The results demonstrate very good agreement with other fountain clocks and previously published optical frequency measurement data at the low $10^{-16}$ level. High duty cycles of the optically stabilized microwave oscillator signal, close to 100\%, have been achieved during weeks of operation. In both fountains, the quartz-based microwave synthesis is only used when maintenance work needs to be done on the frequency comb or the cavity stabilized fiber laser, or in those rare cases when the DRO locking fails.

\section{Conclusion}
\label{sec_conclusion}

The technology of cavity stabilized lasers and femtosecond lasers is well established in many laboratories. Optically stabilized microwave setups as presented are therefore relatively straightforward to realize. The focus of our design is not the ultimate performance in terms of phase noise, but rather reliable continuous operation at a phase noise level which is compatible with fountain clock requirements. Our setup has proven to enable fountain clock frequency instabilities at a level otherwise only accessible by utilizing a cryogenic sapphire oscillator for providing the ultrastable microwave signal \cite{Vian2005}.

We present phase noise measurements of the stabilized microwave oscillator and of the transfer beat frequency signal, and the measured Allan standard deviation of the cavity stabilized fiber laser, locked to the hydrogen maser. We demonstrate a quantum projection noise limited frequency instability of $2.5 \times 10^{-14} (\tau/1\,\mathrm{s})^{-1/2}$ and continuous long-term operation, typically required by fountain clocks. 

Improvements of fountain clock frequency instabilities not only improve the achievable statistical uncertainty in a given measurement time, but also enable more accurate investigations of systematic effects, which in the end may lead to reduced systematic uncertainties.

\section*{Acknowledgment}

The authors would like to thank N.~Huntemann for providing the $^{171}$Yb$^+$ single-ion frequency standard reference signal, G.~Dobrev for his contributions to enhancing the output of the cold atom beam source of CSF2, M.~Kazda for providing the necessary interface in the fountain frequency syntheses to incorporate the optically stabilized microwave signal, and N.~Nemitz for the required fountain control software adaptation.

\ifCLASSOPTIONcaptionsoff
  \newpage
\fi


\bibliographystyle{IEEEtran}
\bibliography{KammMW}

\begin{thebibliography}{10}
\providecommand{\url}[1]{#1}
\csname url@samestyle\endcsname
\providecommand{\newblock}{\relax}
\providecommand{\bibinfo}[2]{#2}
\providecommand{\BIBentrySTDinterwordspacing}{\spaceskip=0pt\relax}
\providecommand{\BIBentryALTinterwordstretchfactor}{4}
\providecommand{\BIBentryALTinterwordspacing}{\spaceskip=\fontdimen2\font plus
\BIBentryALTinterwordstretchfactor\fontdimen3\font minus
  \fontdimen4\font\relax}
\providecommand{\BIBforeignlanguage}[2]{{%
\expandafter\ifx\csname l@#1\endcsname\relax
\typeout{** WARNING: IEEEtran.bst: No hyphenation pattern has been}%
\typeout{** loaded for the language `#1'. Using the pattern for}%
\typeout{** the default language instead.}%
\else
\language=\csname l@#1\endcsname
\fi
#2}}
\providecommand{\BIBdecl}{\relax}
\BIBdecl

\bibitem{Wynands2005}
\BIBentryALTinterwordspacing
R.~Wynands and S.~Weyers, ``{Atomic fountain clocks},'' \emph{Metrologia},
  vol.~42, no.~3, p. S64, 2005. [Online]. Available:
  \url{http://stacks.iop.org/0026-1394/42/i=3/a=S08}
\BIBentrySTDinterwordspacing

\bibitem{Guena2012}
J.~Gu{\'e}na, M.~Abgrall, D.~Rovera, P.~Laurent, B.~Chupin, M.~Lours,
  G.~Santarelli, P.~Rosenbusch, M.~E. Tobar, R.~Li, K.~Gibble, A.~Clairon, and
  S.~Bize, ``{Progress in Atomic Fountains at LNE-SYRTE},'' \emph{IEEE
  Transactions on Ultrasonics, Ferroelectrics, and Frequency Control}, vol.~59,
  no.~3, pp. 391--410, March 2012.

\bibitem{Dobrev2016}
\BIBentryALTinterwordspacing
G.~Dobrev, V.~Gerginov, and S.~Weyers, ``Loading a fountain clock with an
  enhanced low-velocity intense source of atoms,'' \emph{Phys. Rev. A},
  vol.~93, p. 043423, Apr 2016. [Online]. Available:
  \url{http://link.aps.org/doi/10.1103/PhysRevA.93.043423}
\BIBentrySTDinterwordspacing

\bibitem{Santarelli1998}
G.~Santarelli, C.~Audoin, A.~Makdissi, P.~Laurent, G.~Dick, and A.~Clairon,
  ``{Frequency stability degradation of an oscillator slaved to a periodically
  interrogated atomic resonator},'' \emph{IEEE Transactions on Ultrasonics,
  Ferroelectrics, and Frequency Control}, vol.~45, no.~4, pp. 887--894, July
  1998.

\bibitem{Santarelli1999}
\BIBentryALTinterwordspacing
G.~Santarelli, P.~Laurent, P.~Lemonde, A.~Clairon, A.~G. Mann, S.~Chang, A.~N.
  Luiten, and C.~Salomon, ``{Quantum Projection Noise in an Atomic Fountain: A
  High Stability Cesium Frequency Standard},'' \emph{Phys. Rev. Lett.},
  vol.~82, no.~23, pp. 4619--4622, Jun 1999. [Online]. Available:
  \url{http://link.aps.org/doi/10.1103/PhysRevLett.82.4619}
\BIBentrySTDinterwordspacing

\bibitem{Takamizawa2014}
A.~Takamizawa, S.~Yanagimachi, T.~Tanabe, K.~Hagimoto, I.~Hirano, K.~Watabe,
  T.~Ikegami, and J.~G. Hartnett, ``{Atomic Fountain Clock With Very High
  Frequency Stability Employing a Pulse-Tube-Cryocooled Sapphire Oscillator},''
  \emph{IEEE Transactions on Ultrasonics, Ferroelectrics, and Frequency
  Control}, vol.~61, no.~9, pp. 1463--1469, Sept 2014.

\bibitem{Abgrall2016}
M.~Abgrall, J.~Gu\'ena, M.~Lours, G.~Santarelli, M.~E. Tobar, S.~Bize, S.~Grop,
  B.~Dubois, C.~Fluhr, and V.~Giordano, ``{High-Stability Comparison of Atomic
  Fountains Using Two Different Cryogenic Oscillators},'' \emph{IEEE
  Transactions on Ultrasonics, Ferroelectrics, and Frequency Control}, vol.~63,
  no.~8, pp. 1198--1203, Aug 2016.

\bibitem{Puppe2016}
\BIBentryALTinterwordspacing
T.~Puppe, A.~Sell, R.~Kliese, N.~Hoghooghi, A.~Zach, and W.~Kaenders,
  ``{Characterization of a DFG comb showing quadratic scaling of the phase
  noise with frequency},'' \emph{Opt. Lett.}, vol.~41, no.~8, pp. 1877--1880,
  Apr 2016. [Online]. Available:
  \url{http://ol.osa.org/abstract.cfm?URI=ol-41-8-1877}
\BIBentrySTDinterwordspacing

\bibitem{Hudson2005}
\BIBentryALTinterwordspacing
D.~D. Hudson, K.~W. Holman, R.~J. Jones, S.~T. Cundiff, J.~Ye, and D.~J. Jones,
  ``Mode-locked fiber laser frequency-controlled with an intracavity
  electro-optic modulator,'' \emph{Opt. Lett.}, vol.~30, no.~21, pp.
  2948--2950, Nov 2005. [Online]. Available:
  \url{http://ol.osa.org/abstract.cfm?URI=ol-30-21-2948}
\BIBentrySTDinterwordspacing

\bibitem{Zhang2012}
W.~Zhang, M.~Lours, M.~Fischer, R.~Holzwarth, G.~Santarelli, and Y.~{Le Coq},
  ``{Characterizing a Fiber-Based Frequency Comb With Electro-Optic
  Modulator},'' \emph{IEEE Transactions on Ultrasonics, Ferroelectrics, and
  Frequency Control}, vol.~59, no.~3, pp. 432--438, March 2012.

\bibitem{Millo2009}
\BIBentryALTinterwordspacing
J.~Millo, R.~Boudot, M.~Lours, P.~Y. Bourgeois, A.~N. Luiten, Y.~{Le Coq},
  Y.~Kersal\'{e}, and G.~Santarelli, ``Ultra-low-noise microwave extraction
  from fiber-based optical frequency comb,'' \emph{Opt. Lett.}, vol.~34,
  no.~23, pp. 3707--3709, 2009. [Online]. Available:
  \url{http://ol.osa.org/abstract.cfm?URI=ol-34-23-3707}
\BIBentrySTDinterwordspacing

\bibitem{Telle2002}
\BIBentryALTinterwordspacing
H.~Telle, B.~Lipphardt, and J.~Stenger, ``Kerr-lens, mode-locked lasers as
  transfer oscillators for optical frequency measurements,'' \emph{Applied
  Physics B}, vol.~74, no.~1, pp. 1--6, 2002. [Online]. Available:
  \url{http://dx.doi.org/10.1007/s003400100735}
\BIBentrySTDinterwordspacing

\bibitem{Lipphardt2009}
B.~Lipphardt, G.~Grosche, U.~Sterr, C.~Tamm, S.~Weyers, and H.~Schnatz, ``{The
  Stability of an Optical Clock Laser Transferred to the Interrogation
  Oscillator for a Cs Fountain},'' \emph{IEEE Transactions on Instrumentation
  and Measurement}, vol.~58, no.~4, pp. 1258--1262, April 2009.

\bibitem{Weyers2009}
\BIBentryALTinterwordspacing
S.~Weyers, B.~Lipphardt, and H.~Schnatz, ``Reaching the quantum limit in a
  fountain clock using a microwave oscillator phase locked to an ultrastable
  laser,'' \emph{Phys. Rev. A}, vol.~79, no.~3, p. 031803, Mar 2009. [Online].
  Available: \url{http://dx.doi.org/10.1103/PhysRevA.79.031803}
\BIBentrySTDinterwordspacing

\bibitem{Tamm2014}
\BIBentryALTinterwordspacing
C.~Tamm, N.~Huntemann, B.~Lipphardt, V.~Gerginov, N.~Nemitz, M.~Kazda,
  S.~Weyers, and E.~Peik, ``Cs-based optical frequency measurement using
  cross-linked optical and microwave oscillators,'' \emph{Phys. Rev. A},
  vol.~89, p. 023820, Feb 2014. [Online]. Available:
  \url{http://link.aps.org/doi/10.1103/PhysRevA.89.023820}
\BIBentrySTDinterwordspacing

\bibitem{Huntemann2014}
\BIBentryALTinterwordspacing
N.~Huntemann, B.~Lipphardt, C.~Tamm, V.~Gerginov, S.~Weyers, and E.~Peik,
  ``{Improved Limit on a Temporal Variation of ${m}_{p}/{m}_{e}$ from
  Comparisons of ${\mathrm{Yb}}^{+}$ and Cs Atomic Clocks},'' \emph{Phys. Rev.
  Lett.}, vol. 113, p. 210802, Nov 2014. [Online]. Available:
  \url{http://link.aps.org/doi/10.1103/PhysRevLett.113.210802}
\BIBentrySTDinterwordspacing

\bibitem{Jones2000}
\BIBentryALTinterwordspacing
D.~J. Jones, S.~A. Diddams, J.~K. Ranka, A.~Stentz, R.~S. Windeler, J.~L. Hall,
  and S.~T. Cundiff, ``{Carrier-Envelope Phase Control of Femtosecond
  Mode-Locked Lasers and Direct Optical Frequency Synthesis},'' \emph{Science},
  vol. 288, no. 5466, pp. 635--639, 2000. [Online]. Available:
  \url{http://science.sciencemag.org/content/288/5466/635}
\BIBentrySTDinterwordspacing

\bibitem{Holzwarth2000}
\BIBentryALTinterwordspacing
R.~Holzwarth, T.~Udem, T.~W. H\"ansch, J.~C. Knight, W.~J. Wadsworth, and
  P.~S.~J. Russell, ``{Optical Frequency Synthesizer for Precision
  Spectroscopy},'' \emph{Phys. Rev. Lett.}, vol.~85, pp. 2264--2267, Sep 2000.
  [Online]. Available:
  \url{http://link.aps.org/doi/10.1103/PhysRevLett.85.2264}
\BIBentrySTDinterwordspacing

\bibitem{Huntemann2016}
\BIBentryALTinterwordspacing
N.~Huntemann, C.~Sanner, B.~Lipphardt, C.~Tamm, and E.~Peik, ``{Single-Ion
  Atomic Clock with $3\ifmmode\times\else\texttimes\fi{}{10}^{-18}$ Systematic
  Uncertainty},'' \emph{Phys. Rev. Lett.}, vol. 116, p. 063001, Feb 2016.
  [Online]. Available:
  \url{http://link.aps.org/doi/10.1103/PhysRevLett.116.063001}
\BIBentrySTDinterwordspacing

\bibitem{Kramer2004}
G.~Kramer and W.~Klische, ``Extra high precision digital phase recorder,'' in
  \emph{Proceedings of the 18th European Frequency and Time Forum, Guildford},
  2004, pp. 595--602.

\bibitem{Gupta2007}
A.~Sen~Gupta, R.~Schr\"oder, S.~Weyers, and R.~Wynands, ``{A New 9-\textsc{GH}z
  Synthesis Chain for Atomic Fountain Clocks},'' in \emph{Proceedings of the
  2007 Joint Meeting of the European Frequency and Time Forum and the IEEE
  International Frequency Control Symposium}, May/June 2007, pp. 234--237.

\bibitem{Weyers2001}
\BIBentryALTinterwordspacing
S.~Weyers, U.~H\"ubner, R.~Schr\"oder, C.~Tamm, and A.~Bauch, ``{Uncertainty
  evaluation of the atomic caesium fountain CSF1 of the PTB},''
  \emph{Metrologia}, vol.~38, no.~4, p. 343, 2001. [Online]. Available:
  \url{http://stacks.iop.org/0026-1394/38/i=4/a=7}
\BIBentrySTDinterwordspacing

\bibitem{Weyers2001b}
S.~Weyers, A.~Bauch, R.~Schr\"oder, and C.~Tamm, ``{The atomic caesium fountain
  CSF1 of PTB},'' in \emph{{Proceedings of the 6th Symposium on Frequency
  Standards and Metrology}}, P.~Gill, Ed., University of St Andrews, Fife,
  Scotland, 2001, pp. 64--71.

\bibitem{Gerginov2010}
\BIBentryALTinterwordspacing
V.~Gerginov, N.~Nemitz, S.~Weyers, R.~Schr\"oder, D.~Griebsch, and R.~Wynands,
  ``{Uncertainty evaluation of the caesium fountain clock PTB-CSF2},''
  \emph{Metrologia}, vol.~47, no.~1, p.~65, 2010. [Online]. Available:
  \url{http://stacks.iop.org/0026-1394/47/i=1/a=008}
\BIBentrySTDinterwordspacing

\bibitem{Weyers2012}
\BIBentryALTinterwordspacing
S.~Weyers, V.~Gerginov, N.~Nemitz, R.~Li, and K.~Gibble, ``{Distributed cavity
  phase frequency shifts of the caesium fountain PTB-CSF2},''
  \emph{Metrologia}, vol.~49, no.~1, p.~82, 2012. [Online]. Available:
  \url{http://stacks.iop.org/0026-1394/49/i=1/a=012}
\BIBentrySTDinterwordspacing

\bibitem{Taylor2011}
J.~Taylor, S.~Datta, A.~Hati, C.~Nelson, F.~Quinlan, A.~Joshi, and S.~Diddams,
  ``{Characterization of Power-to-Phase Conversion in High-Speed P-I-N
  Photodiodes},'' \emph{IEEE Photonics Journal}, vol.~3, no.~1, pp. 140--151,
  Feb 2011.

\bibitem{Zhang2012a}
\BIBentryALTinterwordspacing
W.~Zhang, T.~Li, M.~Lours, S.~Seidelin, G.~Santarelli, and Y.~Le~Coq,
  ``{Amplitude to phase conversion of InGaAs pin photo-diodes for femtosecond
  lasers microwave signal generation},'' \emph{Applied Physics B}, vol. 106,
  no.~2, pp. 301--308, 2012. [Online]. Available:
  \url{http://dx.doi.org/10.1007/s00340-011-4710-1}
\BIBentrySTDinterwordspacing

\bibitem{Zhang2010}
\BIBentryALTinterwordspacing
W.~Zhang, Z.~Xu, M.~Lours, R.~Boudot, Y.~Kersal\'e, G.~Santarelli, and
  Y.~Le~Coq, ``Sub-100 attoseconds stability optics-to-microwave
  synchronization,'' \emph{Applied Physics Letters}, vol.~96, no.~21, 2010.
  [Online]. Available: \url{http://dx.doi.org/10.1063/1.3431299}
\BIBentrySTDinterwordspacing

\bibitem{Haboucha2011}
\BIBentryALTinterwordspacing
A.~Haboucha, W.~Zhang, T.~Li, M.~Lours, A.~N. Luiten, Y.~{Le Coq}, and
  G.~Santarelli, ``Optical-fiber pulse rate multiplier for ultralow phase-noise
  signal generation,'' \emph{Opt. Lett.}, vol.~36, no.~18, pp. 3654--3656, Sep
  2011. [Online]. Available:
  \url{http://ol.osa.org/abstract.cfm?URI=ol-36-18-3654}
\BIBentrySTDinterwordspacing

\bibitem{Kazda2013}
M.~Kazda, V.~Gerginov, N.~Nemitz, and S.~Weyers, ``{Investigation of Rapid
  Adiabatic Passage for Controlling Collisional Frequency Shifts in a Caesium
  Fountain Clock},'' \emph{IEEE Transactions on Instrumentation and
  Measurement}, vol.~62, no.~10, pp. 2812--2819, Oct 2013.

\bibitem{Vian2005}
C.~Vian, P.~Rosenbusch, H.~Marion, S.~Bize, L.~Cacciapuoti, S.~Zhang,
  M.~Abgrall, D.~Chambon, I.~Maksimovic, P.~Laurent, G.~Santarelli, A.~Clairon,
  A.~Luiten, M.~Tobar, and C.~Salomon, ``{BNM-SYRTE fountains: recent
  results},'' \emph{IEEE Transactions on Instrumentation and Measurement},
  vol.~54, no.~2, pp. 833--836, April 2005.

\end{thebibliography}

\end{document}